\documentstyle[amssymb]{article}

\input{tcilatex}
\begin{document}

\title{Modular Localization, H-Temperatures and the Bethe Ansatz Structure.}
\author{{\bf Bert Schroer} \\
Freie Universit\"at Berlin\\
Institut f\"ur Theoretische Physik}
\date{October 1997\\
Based on talks presented at the conference on ``Noncommutative Geometry and
Applications'', Lissabon Sept.12-15 and at the ``Workshop on Algebraic
QFT'', ESI, Vienna Sept.29-Oct.4}
\maketitle

\begin{abstract}
The recently proposed construction approach to nonperturbative QFT, based on
modular localization, is reviewed and extended. It allows to unify black
holes physics and H-temperatures (H standing for Hawking or Horizon) with
the bootstrap-formfactor program for nonperturbative construction of low
dimensional QFT. In case of on-shell particle number conservation, the
equations characterizing the modular localization spaces for wedges are
Bethe-Ansatz equation in the form as recently obtained in the treatment of
factorizable models.
\end{abstract}

\section{Historical Remarks and Present State}

The modular theory of von Neumann algebras is one of the few theories of
which the foundations received independent contributions from mathematicians
and quantum field theorist; a fact which testifies to the naturalness of the
modular concepts. During the 60's in a tour de force the mathematician
Tomita obtained the most important statements which were received by his
contemporaries with a mixture of surprise and\thinspace disbelief. Shortly
afterwards his Japanese colleague Takesaki \cite{Takesaki} corrected and
further developed the theory\footnote{%
The reader is urged to read the lecture notes of R.Kadison which will be
published in the proceedings of the 1997 Summer School in Portugal on
''Noncommutative Geometry and Applications''.}, this time already using
concepts of the Haag, Hugenholtz, Winnink\cite{Haag}\footnote{%
Whenever references have entered textbooks, we prefer to quote the latter.}
description of the infinite volume (thermodynamic) limit for thermal states
on QFT systems which were elaborated at approximately the same time as
Tomita's contributions. As usual, the physical context of such a conceptual
discovery is somewhat more special. It was the deeper understanding of the
so-called ''KMS property''\footnote{%
Originally just a receipe to avoid the calculation of cumbersome traces in
favour of analyticity properties combined with boundary conditions which was
introduced in the 50's by Kubo, Martin and Schwinger\cite{Haag}}, which
connected the HHW-thermal theory with the modular theory of Tomita and its
improvement by Takesaki \cite{Takesaki} and finally led to what is nowadays
referred to as the Tomita-Takesaki modular theory.

I can only think of one other such {\it natural} ``marriage'' between
physics and mathematics. This is the closely related subfactor theory in von
Neumann algebras of V. Jones \cite{Jones}, in which case the (again much
more special) physical counterpart (the Doplicher-Haag-Roberts theory of
localized endomorphisms \cite{Haag}) preceded the mathematical development
by almost one decade, although the interconnections were only noticed
several years after Vaughn Jones's discovery.

The further development of the T.-T. modular theory (here in the context of
algebraic QFT briefly referred to as ``modular theory'') during the 70's is
characterized by the names A.Connes and H.Araki. These authors extended the
theory by the concept of ``natural cones'' and Connes used the theory for
his famous classification of type III factors \cite{conf}. On the physical
side the development during the 70's is characterized by the beginning of
understanding of the importance of modular theory for the localization
concept in relativistic QFT. Following suggestive ideas of Araki, Eckmann
and Osterwalder \cite{Eck} and later Leyland, Roberts and Testard \cite{Ley}
first recognized the close connection in the context of free fields before
Bisognano and Wichmann finally achieved a more general understanding within
the setting of QFT as formulated by A.S. Wightman \cite{Haag}. During the
80's and 90's there have been many mathematical physics contributions
extending physical and mathematical aspects of modular theory into different
directions. Our own contribution presented in this article is ''inverse'' to
the one of Bisognano and Wichmann \cite{Haag} i.e. it tries to construct
local theories via the concept \cite{S1}\cite{S2} of ''modular
localization'' . As such it has some mathematical aims in common with
investigations of M.Wollenberg \cite{Wo}.

\thinspace In this paper we are interested in the inverse problem: how to
obtain localized states and local algebras (and fields) from modular theory.
Interestingly enough, our more physically motivated approach leads to the 
{\it Main Inverse Problem of QFT}: how to obtain reasonable physical
conditions under which a given admissible scattering operator $S_{s}$
determines uniquely a local QFT. As a kind of side result we obtain new
ideas of how the somewhat elusive (in its nonperturbative aspects) crossing
symmetry together with its associated ''on shell'' analyticity property is
related to the KMS property of the (Rindler) wedge-localized Hawking-Unruh
effect. Our concepts attribute a very fundamental role to the modular
reflection. As such our viewpoint has many things in common with recent work
of Buchholz and Summers \cite{Bu} apart from the fact that they use the
Tomita $J^{\prime }s$ for the reconstruction of space-time properties and a
characterization of a vacuum reference state (or its substitute).

In order to structure our review of some old points in a new light and
combine it with the presentation of new viewpoints, we will follow the
outline in terms of six sections.

\begin{enumerate}
\item  {\bf Historical Remarks and Present State}

\item  {\bf Liberation from ``Field Coordinates''}

\item  {\bf \ H-temperature (Hawking, Unruh, Sewell, Wald, Kay and others)}

\item  {\bf \ Modular Localization and Factorizing Theories..}

\item  {\bf \ General Interactions and Modular Localization.}

\item  {\bf \ Present Conclusions, Outlook and possible Connection with
other Ideas.}
\end{enumerate}

The second section describes the path from Wigner's positive energy
representations to modular localization subspaces and local nets and avoids
the use of ``field coordinates'' altogether \cite{S1}. The localization
peculiarities of massless theories with helicities $h\geq 1$ (which
constitute the physical origin of the ''gauge'' phenomenon) as well as those
of ``continuous'' helicity are briefly mentioned and the necessity for
noncompact modular localization of anyons is explained.

The third section strengthens the field theoretic interpretation of thermal
aspect of the Hawking-Unruh effect as consequences of localization. The big
Latin letter $H$ has not been chosen as a result of recent fashions (hiding
unfamiliar and insufficiently defined inventions behind familiar sounding
letters), but rather represents the intended double meaning of either
Hawking or (bifurcated) Horizon depending on the context. In the present
context this temperature may also be called ``modular localization
temperature'' and it is associated with its own ``temperature Hilbert
space'' as will be justified in section 3.

The fourth section contains the use of modular localization for the
description and explicit construction of interacting theories in d=1+1
dimensions. In these notes the computational successful, but insufficiently
understood bootstrap-formfactor approach of Karowski and Weiss \cite{KW} ,
as well as of Smirnovs extensions \cite{Smi}, and in particular the more
recent contributions of Babujian, Fring, Karowski and Zapletal \cite{BKZ}%
\cite{BFK} will be newly interpreted, modified and extended in the light of
modular localization \cite{S2}. The surprising new result is that the
Riemann-Hilbert problem and the Bethe Ansatz method, which appears in the
work of the last 4 authors is identical to the properties of the modular
localization equations for wedges in relativistic QFT with on shell particle
conservation (but arbitrarily complicated off shell nonconservation
behavior).

The fourth section contains remarks on higher dimensional systems. In
particular for d=1+3 theories, for which on shell conservation of particle
number is known to be incompatible with interactions, our remarks are
presently very speculative and preliminary indeed.

Finally we use the fifth section to draw some general conclusions about the
structure of QFT and compare the modular localization concept to attempts
which base quantum physics on more global concepts (string theory, method of
noncommutative geometry for gravity and electro-weak interactions).

\section{Liberation from Free Field Co-ordinates}

As explained elsewhere \cite{S1}, one may use the Wigner representation
theory for positive energy representations in order to construct fields from
particle states. For $d=3+1$ space-time dimensions there are two families of
representation: $(m,s)$ and $(0,h)$. Here $m$ is the mass and designates
massive representation and $s$ and $h$ are the spins resp. the helicites $h$%
. These are invariants of the representations (``Casimirs'') which refer to
the Wigner ``little'' group; in the first case to SU(2) in which case $s=$
(half) integer, and for $m=0$ to the little group (fixed point group of a
momentum $\neq 0$ on the light cone) $\tilde{E}(2)$ which is the two-fold
covering of the euclidean group in the plane. The zero mass representations
split into two families. For the ``neutrino-photon family'' the little group
has a nonfaithful representation (the ``translative'' part is trivially
represented) whereas for so-called ``continuous $h$ representation'' the
representation is faithful but allows no identification with known zero mass
particles.

In the massive case, the transition to covariant fields is most conveniently
done with the help of intertwiners between the Wigner spin $s$
representations $D^{(s)}(R(\Lambda ,p))$ which involve the $\Lambda ,p$
dependent Wigner rotation $R$ and the finite dimensional covariant
representation of the Lorentz-group $D^{[A,B]}$ 
\begin{equation}
u(p)D^{(s)}(R(\Lambda ,p))=D^{[A,B]}(\Lambda )u(\Lambda ^{-1}p)  \label{1}
\end{equation}
The only restriction is: 
\begin{equation}
\mid A-B\mid \le s\le A+B  \label{2}
\end{equation}
which leaves infinitely many $A,B$ (half integer) choices for a given $s$.
Here the $u(p)$ intertwiner is a rectangular matrix consisting of $2s+1$
column vectors $u(p,s_{3}),s_{3}=-s,...,+s$ of length $(2A+1)(2B+1)$. Its
explicit construction using Clebsch-Gordan methods can be found in
Weinberg's book \cite{Wei}.Analogously there exist antiparticle (opposite
charge) $v(p)$ intertwiners: $D^{(s)*}(R(\Lambda ,p)\longrightarrow
D^{[A,B]}(\Lambda )$. The covariant field is then of the form: 
\begin{equation}
\psi ^{[A,B]}(x)=\frac{1}{(2\pi )^{3/2}}\int
(e^{-ipx}\sum_{s_{3}}u(p_{1}s_{3})a(p_{1}s_{3})+e^{ipx}%
\sum_{s_{s}}v(p_{1}s_{3})b^{*}(p_{1}s_{3}))\frac{d^{3}p}{2\omega }.
\label{int}
\end{equation}
where $a(p)$ and $b^{*}(p)$ are annihilation (creation) operators in a
Fockspace for particles (antiparticles). The (only at first sight, as it
fortunately turns out) bad news is that in we lost the Wigner unicity: there
are now infinitely many $\psi ^{[A,B]}$ fields with varying $A,B$ but all
belonging to the same $(m,s)$-Wigner representation and living in the same
Fock space. Only one of these fields is ``Eulerian'' (examples: for $s=\frac{%
1}{2}$ Dirac, for $s=\frac{3}{2}$ Rarita-Schwinger) i.e. the transformation
property of $\psi $ is a consequence of the nature of a linear field
equation and which is derivable by an action principle from a Lagrangian.
Non-Eulerian fields as e.g. Weinberg's $D^{\left[ j,0\right] }+D^{\left[
0,j\right] }$ fields for $j>\frac{3}{2},$ cannot be used in a canonical
quantization scheme or in a formalism of functional integration because the
corresponding field equations have more solutions than allowed by the
physical degrees of freedom ( in fact they have tachyonic solutions). The
use of formula (\ref{int}) with the correct $u,v$ intertwiners in the (on
shell) Bogoliubov-Shirkov approach based on causality is however legitimate.
Naturally from the point of view of the Wigner theory which is totally
intrinsic and does not use quantization ideas, there is no preference of
Eulerian versus non Eulerian fields.

It turns out that the above family of fields corresponding to $(m,s)$
constitute the linear part of the associated ``Borchers class'' \cite{Haag}.
For bosonic fields the latter is defined as: 
\begin{equation}
B(\psi )=\left\{ \chi (s)\mid [\chi (x),\psi (y)]=0,\,\,\,(x-y)^{2}<0\right\}
\label{4}
\end{equation}
If we only consider cyclic (with respect to the vacuum) relatively local
fields, than we obtain transitivity in addition to the auto-locality of the
resulting fields. This class depends only on $\left( m,s\right) $ and is
generated by the Wick-monomials of $\psi $. A mathematically and
conceptually more managable object which is manifestly independent of the
chosen (m,s) Fock-space field, is the local von Neumann algebra generated by 
$\psi $: 
\begin{equation}
{\cal O}\rightarrow {\cal A}({\cal O},\psi )={\cal A}({\cal O},\chi )
\end{equation}
Here $\chi \sim \psi $ is any cyclic (locally equivalent) field in the same
Borchers class of $\psi .$

Now we have reached our first goal: the lack of uniqueness of local $(m,s)$
fields is explained in terms of the arbitrariness in the choice of ``field
coordinates'' which generate the same net of (observable) von Neumann
algebras. According to the physical interpretation in algebraic QFT this
means that the physics does not depend on the concretely chosen (cyclic)
field.

Since algebraic QFT shuns inventions and favors discoveries, it is deeply
satisfying that there are arguments that every causal net fulfilling certain
spectral properties is automatically ``coordinatizable''. For chiral
conformal theories there exists even a rigorous proof \cite{Joe}. So one can
be reasonably sure that the physical content has not been changed as
compared to the standard Wightman approach. The use of local field
coordinates tends to make geometric localization properties of the algebras
manifest. But only if there exist pointlike covariant generators which
create charged states (counter example: for Maxwellian charges they do not
exist) the localization can be encoded into classical smearing function. The
localization concept is ``maximally classical'' for the free Weyl and CAR
algebras which are just function algebras with a noncommutative product
structure. For these special cases the differential geometric concepts as
fibre bundles may be directly used in local quantum physics. Outside of
these special context, the only reliable methods are the von Neumann algebra
methods of algebraic QFT. In that case the quantum localization may deviate
from the classical geometric concepts and use of field coordinates is less
useful.

In the following we describe a way to construct the interaction-free nets
directly thus bypassing the use of field coordinates alltogether. We use the
d=3+1 Wigner (m,s)-representations as an illustrative example. In case of
charged particles (particles$\neq $antiparticles) we double the Wigner
representation space: 
\begin{equation}
H=H_{Wig}^{p}\oplus H_{Wig}^{\bar{p}}
\end{equation}
in order to incorporate the charge conjugation operation as an (antilinear
in the Wigner theory) operator involving the p-\={p}-flip. On this extended
Wigner space one can represent the full Poincar\'{e} group where those
reflections which change the direction of time are antiunitarily
represented. For the modular localization in a wedge we only need the
standard L-boost $\Lambda (\chi )$ and the standard reflection $r$ which (by
definition) are associated with the $t-x$ wedge: 
\begin{equation}
\delta ^{i\tau }\equiv \pi _{Wig}(\Lambda (\chi =2\pi \tau ))
\end{equation}
\begin{equation}
j\equiv \pi _{Wig}(r)
\end{equation}
These operators have a simple action on the p-space (possibly) doubled
Wigner wave functions, in particular: 
\begin{equation}
(j\psi )(p)\simeq \left( 
\begin{array}{ll}
0 & -1 \\ 
1 & 0
\end{array}
\right) \bar{\psi}(p_{0},p_{1},-p_{2},-p_{3})
\end{equation}
By functional calculus we form $\delta ^{\frac{1}{2}}$ and define: 
\begin{equation}
s\equiv j\delta ^{\frac{1}{2}}
\end{equation}
This unbounded antilinear densely defined operator $s$ is involutive on its
domain: $s^{2}=1.$ Its -1 eigenspace is a real closed subspace $H_{R}$ of $H$
which allows the following characterization of the domain of $s:$%
\begin{eqnarray}
dom(s) &=&H_{R}+iH_{R} \\
s(h_{1}+ih_{2}) &=&-h_{1}+ih_{2}  \nonumber
\end{eqnarray}
Defining: 
\begin{equation}
H_{R}(W)\equiv U(g)H_{R},\,\,\,W=gW_{stand}
\end{equation}
where g is an appropriate Poincar\'{e} transformation, we find the following
theorem:

\begin{theorem}
(Brunetti, Guido and Longo, in preparation): $H_{R}(W)$ is a net of real
Hilbert spaces i.e. $H_{R}(W_{1})\subsetneq H_{R}(W_{2})$ if $%
W_{1}\subsetneq W_{2}$.
\end{theorem}

Although we will not give the proof \cite{BGL}, it turns out to be quite
easy, at least if one is familiar with the work of Borchers \cite{Bo2} which
already contains the idea on how positive energy translations are related
with compressions of algebras.

If we now define: 
\begin{equation}
H_{R}({\cal O})\equiv \bigcap_{W\supset {\cal O}}H_{R}(W)
\end{equation}
Then it is easily seen (even without the use of the u,v-intertwiners) that
the spaces $H_{R}({\cal O})+iH_{R}({\cal O})$ are still dense in $H_{Wig}$
and that the formula: 
\begin{equation}
s({\cal O})(h_{1}+ih_{2})\equiv -h_{1}+ih_{2}
\end{equation}
defines a closed involutive operator with a polar decomposition: 
\begin{equation}
s({\cal O})=j({\cal O})\delta ({\cal O})^{\frac{1}{2}}
\end{equation}
Although now $j(O)$ and $\delta ({\cal O})^{i\tau }$ have no obvious
geometric interpretation, there is still a bit of geometry left, as the
following theorem shows:

\begin{theorem}
The $H_{R}({\cal O})$ form an orthocomplemented net of closed real Hilbert
spaces, i.e. the following ''duality'' holds: $H_{R}({\cal O}^{\prime
})=H_{R}({\cal O})^{\prime }=iH_{R}^{\bot }({\cal O}).$
\end{theorem}

Here ${\cal O}^{\prime }$ denotes the causal complement, $H_{R}^{\bot }$ the
real orthogonal complement in the sense of the inner product $Re\left( \psi
,\varphi \right) $ and $H_{R}^{\prime }$ is the symplectic complement in the
sense of $\func{Im}\left( \psi ,\varphi \right) .$

The direct construction of the interaction-free algebraic bosonic net for
(m,s=integer) is now achieved by converting the ''premodular'' theory of
real subspaces of the Wigner space into the Tomita-Takesaki modular theory
for nets of von Neumann algebras using the Weyl functor:

\begin{theorem}
The application of the Weyl functor ${\cal F}$ to the net of real spaces: 
\begin{equation}
H_{R}({\cal O})\stackrel{\cal F}{\rightarrow }{\cal A}({\cal O})\equiv
alg\left\{ W(f)\left| f\in H_{R}({\cal O})\right. \right\} 
\end{equation}
leads to a net of von Neumann algebras in ${\cal H}_{Fock}\,$which are in
``standard position'' with respect to the vacuum state with a modular theory
which, if restricted to the Fock vacuum $\Omega ,$ is geometric: 
\begin{eqnarray}
{\cal F}(s) &=&S,\,\,\,SA\Omega =A^{*}\Omega ,\,\,\,A\in {\cal A}(W) \\
S &=&J\Delta ^{\frac{1}{2}},\,\,\,J={\cal F}(j),\,\,\,\Delta ^{i\tau }={\cal %
F}(\delta ^{i\tau })  \nonumber
\end{eqnarray}
The proof of this theorem uses the functorial formalism of \cite{Ley}
\end{theorem}

Clearly the $W$ or ${\cal O}$ indexing of the spaces corresponds to a
localization concept via modular theory. Specifically $H_{R}({\cal O}%
)+iH_{R}({\cal O})$ is a certain closure of the one particle component of
the Reeh-Schlieder domain belonging to the localization region ${\cal O}.$
Although for general localization region the modular operators are not
geometric, there is one remaining geometric statement which presents itself
in the form of an algebraic duality property: 
\begin{equation}
{\cal A}({\cal O}^{\prime })={\cal A}({\cal O})^{\prime
},\,\,\,\,Haag\,\,Duality
\end{equation}
Here the prime on the von Neumann algebra has the standard meaning of
commutant.

It is very instructive to write the modular wedge localization equations in
Fock space more explicitly.. For simplicity we consider the case of d=1+1
free fields because in that case n-particle states can be characterized
solely in terms of rapidities $\theta $ (there is only one wedge and its
opposite). We find: 
\begin{equation}
S\psi =\psi \Leftrightarrow \overline{f_{n}(\theta _{1},.....\theta _{n})}%
=f_{n}(\theta _{1}+i\pi ,....\theta _{n}+i\pi )  \label{free}
\end{equation}
where $f_{n}$ are the n-particle component wave functions in momentum space
(rapidities). This is a boundary relation for an analytic function which is
analytic inside the multidimensional $i\pi $ strip; the analyticity beeing a
consequence of the fact that $\psi $ must be contained in the domain of $%
\Delta ^{\frac{1}{2}}$ because $domS=dom\Delta ^{\frac{1}{2}}.$ The analytic
``master'' function $f_{n}$ has different real boundaries corresponding to
the permutations of the standard ordering $\theta _{1}\leq \theta _{2}\leq
...\leq \theta _{n}.$ As for Wightman functions in x-space, the different
orders are obtained by different ways of letting the imaginary parts
approach zero.

Whereas in Wigner space the subspace belonging to e.g. a double cone is
simply the intersection of the two wedges which define this double cone (at
least in case of (half)integer L-spin), these net properties are not true in
Fock space. So in the interacting case we cannot expect the existence of a
functor from real subspaces of Fock space to local algebras.

In the following we make some schematic additions and completions which
highlight the modular localization concept for more general cases.

\begin{itemize}
\item  (1) In the case of $m\neq 0,s=$ halfinteger, the Wigner theory
produces a mismatch between the ``quantum'' and the ``geometric'' opposite
of $H_{R}(W),$ which however is easily taken care of by an additional factor 
$i$ (interchange of symplectic complement with real orthogonal complement).
This (via the physical localization property) requires the application of
the CAR-functor instead of the CCR-functor as well as the introduction of
the well-known Klein transformation $K$ which takes care of the above
mismatch in Fockspace: 
\begin{eqnarray}
J &=&K{\cal F}_{CAR}(ij)K^{-1} \\
{\cal A}({\cal O}^{\prime }) &=&K{\cal A}({\cal O})^{\prime
}K^{-1},\,\,\,twisted\,\,Haag\,\,\,Duality  \nonumber
\end{eqnarray}

\item  (2) For $m=0,h=$(half)integer, as a consequence of the nonfaithful
representation of the zero mass little group $E(2)$ (the two-dimensional
euclidean group or rather its two-fold covering), the set of possible u-v
intertwiners is limited by the selection rule: $\left| A-B\right| =\pm h.$
This means on the one hand that there are no covariant intertwiners which
lead to $D^{\left[ \frac{1}{2},\frac{1}{2}\right] }$ (vector-potential of
classical Maxwell theory), $D^{\left[ \frac{1}{2},\frac{1}{2}\right]
}\otimes (D^{\left[ \frac{1}{2},0\right] }+D^{\left[ 0,\frac{1}{2}\right] })$
(Rarita-Schwinger potential for massless particles), gravitational
potentials etc. On the other hand, all local bilinear expressions in the
allowed covariant intertwiners vanish and hence cannot be used in order to
rewrite the Wigner inner product (for e.g. $h=1$ in terms of field strength
intertwiners $F_{\mu \nu }(p)).$ A reasonable compromise consists in
relaxing on strict L-covariance and compact (double cone) modular
localization but retaining the relation with the Wigner inner product. One
then may describe the Wigner space in terms of polarization vector dependent
vector-potentials on the light cone which have the following affine Lorentz
transformation: 
\begin{eqnarray}
\left( U(\Lambda )A\right) _{\mu }(p,e) &=&\Lambda _{\mu }^{\nu }A_{\nu
}(\Lambda ^{-1}p,\Lambda ^{-1}e) \\
&=&\Lambda _{\mu }^{\nu }A_{\nu }(\Lambda ^{-1}p,e)+p_{\mu }G(p,\Lambda ,e) 
\nonumber
\end{eqnarray}
where the ''gauge'' contribution $G$ by which one has to re-gauge in order
to refer to the original spacelike polarization vector $e$ is a nonlocal
term which follows from the above definitions. This description is the only
one if one follows the above logic of implementing the modular localization
for the $(0,h=1)$ Wigner representation. After applying the Weyl functor, we
obtain a local covariant net theory which is described in terms of slightly
nonlocal semiinfinite stringlike field coordinates whose relation to the
local $F_{\mu \nu }(x)$ field strengths is given by: 
\begin{equation}
A_{\mu }(x,e)=\int_{0}^{\infty }e^{\nu }F_{\mu \nu }(x-es)ds
\end{equation}
If we now define the modular localization subspaces as before by starting
from the wedge region, we find that the (smoothened versions of) the
vectorpotentials are members of these subspaces (or their translates) as
long as the spacelike directions $e$ point inside the wedges. They are lost
if we form the localization spaces belonging to e.g. double cone regions
regions. Hence these stringlike localized vector potentials appear in a
natural way in our modular localization approach for the wedge regions.
Whereas the natural use of such nonpointlike objects in a future interacting
theory based on modular localization may be possible, in the present
formulation of gauge theories they have not been used. There, one aims for a
description in which the affine contribution of the L-boosts is absent.
Adapting the Kostant-Sternberg analysis of constraint symplectic geometry 
\cite{Du}, one can canonically derive the Kugo-Ojima \cite{Ojima} operator
version of the Faddeev-Popov-BRS description of local free photon fields.
Preliminary work by G. Scharf, T. Hurt, M Duetsch, F. Krahe, K. Fredenhagen
and R. Stora \cite{SHDK} point to the conjecture that perturbative gauge
theories can be formulated as deformatios of the free photon situation
within the Bogoliubov Shirkov Epstein Glaser \cite{EG} framework. Whereas
such an operator formulation may be considered as progress as compared to
the functional calculus of Faddev-Popov (operators are much closer to local
quantum physics than euclidean functional integrals) because it brings the
idea underlying gauge theory a bit closer to Wigner's classification theory
of particles, one should not use the BRS formalism as \cite{BRS} a soft
cushion to rest, since (in my opinion) then one would lose the tremendous
enigmatic power which still resides in these half-solved physical gauge
problems of finding a conceptually acceptable physical description of
interactions involving vector fields with weaker localization properties. It
may very well be that this last step\footnote{%
The step of avoiding the introduction of ghosts which in a later stage will
be eliminated again.} is only possible by using modular localization ideas
in addition to perturbative deformations.

\item  (3) $d=3+1,m=0,h$ ``continuous'' ; $d=2+1,m\neq 0,s\neq $%
(half)integer.
\end{itemize}

The common feature of these cases is that they do not admit compact
localization i.e. $H_{R}({\cal O})=\left\{ 0\right\} ,$ but $H_{R}(W_{1}\cap
W_{2})\neq \left\{ 0\right\} $ for $W_{1}\cap W_{2}$ $\neq \varnothing .$ In
particular for d=1+2 this means that the spacelike cones have nontrivial
localization spaces. However the attempt to construct a local net of spaces
including the spacelike cone regions fails \cite{Mund} since the
intersection of spaces turns out to be genuinely bigger than the
localization space belonging to the intersected region. This fits nicely
together with observations about scattering theory showing that the
multiparticle in-spaces for plektons (including anyons) are not tensor
product of Wigner spaces \cite{Fre}.There is another theorem which shows
that a field which obeys free field equations cannot fulfill anyonic
statistics \cite{Stei}\cite{Mund}, in other words the two-point functions of
anyonic operators have off shell creation even if the asymptotic particle
number is conserved. Such situations, similar to the d=1+1 factorizing
theories, cannot be unraveled by Wigner's theory alone; also the structure
of asymptotic multiparticle states of scattering theory is needed.
Concerning the d=1+2 continuous helicity situation, the a priori best
localization and the ``freest'' field theory behind it still needs to be
investigated. The obligation of a theoretical physicist is not to refer to
``nature not making use of these representations''\footnote{%
Just imagine how an application of this argument would have influenced the
course of supersymmetry in the last twenty years.} but rather to argue that
some localization aspects, which for ordinary particles are required in
addition to the irreducibility and positive energy, could be possibly
missing. But of course there are also particle-like objects as quarks which
are not ordinary particles in the sense of Wigner+compact localization..

\section{ H-Temperature and Modular Localization}

In modular theory the dense set of vectors which are obtained by applying
(local) von Neumann algebras in standard position to the standard (vacuum)
vector forms a core for the Tomita operator $S.$ The domain of $S$ can then
be described in terms of the +1 closed real subspace of $S.$ In terms of the
``premodular'' objects $s$ in Wigner space and the modular Tomita operators $%
S$ in Fock space we introduce the following nets of wedge-localized dense
subspaces: 
\begin{equation}
H_{R}(W)+iH_{R}(W)=dom(s)\subset H_{Wigner}
\end{equation}

\begin{equation}
{\cal H}_{R}(W)+i{\cal H}_{R}(W)=dom(S)\subset {\cal H}_{Fock}
\end{equation}

These dense subspaces become Hilbertspaces in their own right if we use the
graph norm of the Tomita operators. For the $s$-operators in Wigner space we
have: 
\begin{equation}
\left( f,g\right) _{Wigner}\rightarrow \left( f,g\right) _{G}=\left(
f,g\right) _{Wigner}+\overline{\left( sf,sg\right) }_{Wigner}
\end{equation}
The graph topology insures that the wave functions are strip analytic in the
wedge rapidity: $p_{0}=m(p_{\perp })\cosh \theta ,\,\,\,\,p_{1}=m(p_{\perp
})\sinh \theta ,\,\,\,m(p_{\perp })=\sqrt{m^{2}+p_{\perp }^{2}.}$ This is
precisely the analyticity prerequisite for the validity of the KMS property.
Let us look at the thermal localization properties of charged scalar Bosons.
For $f,g\in H_{R}(W)\in H_{Wig}$ we find: 
\begin{eqnarray}
\left( f,g\right) _{Wigner}^{W} &=&\left\langle A(\overline{\hat{f}})A^{*}(%
\hat{g})\right\rangle _{thermal}\stackrel{KMS}{=}\left\langle A^{*}(\hat{g}%
)\Delta A(\hat{f})\right\rangle _{thermal} \\
&&\stackrel{CCR}{=}-\left[ A(\hat{f}),A^{*}(\delta \hat{g})\right] +\left(
f,\delta g\right) _{Wig}^{W},\,\,\,\,\delta =e^{2\pi K}
\end{eqnarray}
\begin{eqnarray}
&\curvearrowright &\left( f,g\right) _{Wig}^{W}=-\left[ A(\hat{f}),A^{*}(%
\frac{\delta }{1-\delta }\hat{g})\right] \\
&=&-\left( f,\frac{\delta }{1-\delta }g\right) _{Wig,p}+\left( \frac{\delta 
}{1-\delta }\delta ^{\frac{1}{2}}\bar{g},\delta ^{\frac{1}{2}}\bar{f}\right)
_{Wig,\bar{p}}  \nonumber
\end{eqnarray}
Here we used smeared fields in intermediate steps. The x-space wedge
supported smearing functions $\hat{f},\hat{g}$ have the Wigner momentum
space wave functions $f,g$ as their on shell restrictions. At the end we
eliminate the field commutator in terms of the Wigner theory of particles 
{\it and antiparticles }whereby the the restriction to the wedge is done
automatically via the domain requirements in the antiparticle term.

So the temperature dependence of localized states becomes manifest and the
difference of a localization- and a heat bath- temperature shows up in the
difference between the two sided-spectrum of the Lorentz-boost generator $K$
and the one-sided spectrum of the hamiltonian $H$ which results in the
unboundedness of $\delta $ in the first case. The fact that the standard
boost $K\,$appears instead of the hamiltonian leads to somewhat different
energy distribution functions. In particular one is advised to discuss
matters of statistics not in Fourier space but rather in the space where
they belong, namely spacetime. For those readers who are familiar with
Unruh's work we mention that the Unruh hamiltonian is different from $2{\cal %
\pi }$ by a factor $\frac{1}{a}$ where a is the acceleration.

The very special free field formalism may be generalized into two directions:

\begin{itemize}
\item  (a) interacting fields

\item  (b) curved spacetime
\end{itemize}

For low-dimensional theories (a) will be discussed in the next section. For
the generalization (b) to curved space time (e.g. the Schwarzschild
solution) it turns out that only the existence of a bifurcated horizon
together with a certain behaviour near that horizon matters (``surface
gravitation'') \cite{Sewell}. In the standard treatment one needs isometries
in spacetime. The idea of modular localization suggests to consider also
e.g. double cones for which there is no spacetime isometry but only an
isometry in $H_{Wigner}$ or ${\cal H}_{Fock}.$ Of course such enlargements
of spaces in order to have a better formulation (or even a solution) of a
problem are a commonplace in modern mathematics, particularly in
noncommutative geometry\footnote{%
Here the unforgettable Gunnar K\"{a}llen comes to my mind who used to call
tricks like this ``Methode Erlk\"{o}nig'' which refers to a famous poem of
Goethe which contains the line ``...und bist du nicht willig so brauch ich
Gewalt...''.}. The idea is that one enlarges the isometries by geometrical
``fuzzy'' ones which only near the horizon may loose their spacetime
fuzziness.

In this context it would be very important to understand the (nongeometric)
modular theory of e.g. the double cone algebra of a massive free field. Fom
the folium of states one may want to select that vector, with respect to
which the algebra has a least fuzzy (most geometric) behaviour under the
action of the modular group. Appealing to the net subtended by spheres at
time t=0 one realizes that algebras localized in these spheres are
independent of the mass. Since m=0 leads to a geometric modular situation%
\footnote{%
The modular group is a one-parametric subgroup of the conformal group.} for
the pair (${\cal A}_{m=0}(S),\left| 0\right\rangle _{m=0},$ and since the
nonlocality of the massive theory in the subtended double cones is only the
result of the fuzzy propagation inside the light cone (the breakdown of
Huygens principle or the ``reverberation'' phenomenon), the fuzziness of the
modular group for the pair (${\cal A}_{m\neq 0}(C(S)),$ $\left|
0\right\rangle _{m\neq 0})$ is a pure propagation phenomenon i.e. can be
understood in terms of the deviation from Huygens principle. In view of the
recent micro-local spectrom condition one expects such nonlocal cases to
have modular groups whose generators are pseudo-differential instead of
(local) differential operators \cite{Fred}.

The Hilbert space setting of modular localization offers also a deeper
physical understanding of the universal domain ${\cal D}$ which plays a
rather technical role in the Wightman framework In the modular localization
approach the necessity for such a domain appears if one wants to come from
the net of localization spaces which receive their natural topology from the
(graphs) net of Tomita operators $\bar{S}({\cal O})$ to a net of (unbounded)
polynomial algebras ${\cal P}({\cal O})$ such that: 
\begin{equation}
dom\,\,\bar{S}({\cal O})\cap {\cal D}={\cal P}({\cal O})\Omega =dom\,\,{\cal %
P}({\cal O})
\end{equation}
this domain is of course also expected to be equal to ${\cal A}({\cal O}%
)\Omega .$ Here we used a more precise notation which distinguishes between
the operator $S$ defined on the core ${\cal A}(0)\Omega $ and its closure $%
\bar{S}$ which is defined on ${\cal H}_{R}({\cal O})+i{\cal H}_{R}({\cal O}%
). $

\section{Modular Localization and Factorizing Theories}

{\it \ ``In diesem Fall und ueberhaupt, kommt es ganz anders als man
glaubt''. (W.Busch)}

{\it [In this special case, as almost always, things happen completely
different to expectations.]}

In this section we will show that the ranges of spaces obtained by applying
all ${\cal O}$-localized (Wightman) fields (or the operator algebra ${\cal %
A(O)}$) to the vacuum, can be used for the nonperturbative construction of
QFT's. This somewhat unexpected state of affairs comes about through modular
localization. Although the modular localization concept is a general
structural property of QFT\footnote{%
Its remoteness from perturbative structures may be the reason why it was
only discovered rather late.}, its constructive use is presently limited to
factorizable (integrable) QFT models. We remind the reader that
``factorizable'' in the intrinsic physical interpretation of algebraic QFT
means ``long-distance representative'' (in the sense of the S-matrix) in a
given superselection class, in other words each general d=1+1 theory has an
asymptotic companion which has the same supersection sectors ($\simeq $ same
particle structure or incoming Fock space) but vastly simplified dynamics
associated to a factorizing S-matrix \cite{S2}. In this paper our goal will
be limited to the modular interpretation and basic field theoretic
understanding of the bootstrap-formfactor program which presently is largely
a collection of plausible cooking recipes \cite{Smi}. Its computational
power e.g. for enlarging the class of soluble models will be shown in
separate future work.

All applications of modular localization to interacting theories are based
on the observation that in asymptotically complete theories with a mass gap,
the full interaction resides in the Tomita operator $J(W),$ whereas the
modular group $\Delta ^{i\tau }(W)$ for wedges (being equal to Lorentz
boosts) is blind against interactions (the representation of the
Poincar\`{e} group is already defined on the free incoming states). In other
words the interaction resides in those disconnected parts of the
Poincar\'{e} group which involve antiunitary time reflections. The Haag
Ruelle scattering theory together with the asymptotic completeness easily
yield (for each wedge): 
\begin{equation}
J=S_{s}J_{0},\,\,\,\,\Delta ^{i\tau }=\Delta _{0}^{i\tau }
\end{equation}
where the subscript 0 refers to the free incoming situation and we have
omitted the reference to the particular wedge. The most convenient form for
this equation is: 
\begin{equation}
S=S_{s}S_{0}
\end{equation}
where $S$ and $S_{0}$ are the antiunitary Tomita operators and $S_{s}$ is
the scattering operator. Therefore the scattering operator $S_{s}$ in
relativistic QFT has two interpretations: it is a global operator in the
sense of large time limits and a modular localization interpretation of
measuring the deviation of $J$ or $S$ from their free field values. This
modular aspect is characteristic of local quantum physics and has no
counterpart in nonrelativistic theory or quantum mechanics. The modular
subspace of ${\cal H}_{Fock}={\cal H}_{in\text{ }}$for the standard wedge: 
\begin{eqnarray}
S_{s}S_{0}{\cal H}_{R} &=&{\cal H}_{R} \\
S_{s}S_{0}\psi &=&-\psi ,\,\,\,\psi \in {\cal H}_{R}
\end{eqnarray}
for general S-matrix is a rather unmanageable object. However for scattering
matrices $S_{s}$ which commute with the incoming particle number and have
the Yang-Baxter structure it will be shown that these equations take on the
form of Bethe-Ansatz equations which can be solved by the (nested)
Bethe-Ansatz method. Before we explain this we will look at the simplest of
such d=1+1 models which is the Federbush model. The model is so simple that
it can be solved by any field theoretic method including the Lagrangian
method. The model consists in coupling two species of Dirac fermions via a
(parity violating) current-pseudocurrent coupling \cite{Wigh}\cite{SW}: 
\begin{equation}
{\cal L}_{int}=g:j_{\mu }^{I}j_{\nu }^{II}:\varepsilon ^{\mu \nu
},\,\,\,\,j_{\mu }=:\bar{\psi}\gamma _{\mu }\psi :
\end{equation}
One easily verifies that: 
\begin{eqnarray}
\psi _{I}(x) &=&\psi _{I}^{(0)}(x)\vdots e^{ig\Phi _{II}^{(l)}(x)}\vdots
\label{local} \\
\psi _{II}(x) &=&\psi _{II}^{(0)}(x)\vdots e^{ig\Phi _{I}^{(r)}(x)}\vdots 
\nonumber
\end{eqnarray}
where $\Phi ^{(l,r)}=\int_{x^{\prime }\lessgtr x}j_{0}dx^{\prime }$ is a
potential of $j_{\mu 5}$ i.e. $\partial _{\mu }\Phi \sim \varepsilon _{\mu
\nu }j^{\nu }=j_{\mu 5}$ and the superscript l,r refers to whether we choose
the integration region for the line integral on the spacelike left or right
of x. The triple ordering is needed in order to keep the closest connection
with classical geometry and localization and in particular to maintain the
validity of the field equation in the quantum theory; for its meaning we
refer to the above papers. This conceptually simpler triple ordering can be
recast into the form of the analytically (computational) simpler standard
Fermion Wick-ordering. Although in this latter description the classical
locality is lost, the quantum exponential do still define local
Fermi-fields; in the case of relative commutation of $\psi _{I}$ with $\psi
_{II}$ the contributions from the exponential (disorder fields) compensate.
Despite the involved looking local fields (\ref{local}), the wedge algebras
are of utmostl simplicity: 
\begin{eqnarray}
{\cal A}(W) &=&alg\left\{ \psi _{I}^{(0)}(f)U_{II}(g),\psi
_{II}^{(0)}(h);suppf,h\in W\right\}  \label{alg} \\
{\cal A}(W^{\prime }) &=&{\cal A}(W)_{Klein}^{^{\prime }}=alg\left\{ \psi
_{I}^{(0)}(f),\psi _{II}^{(0)}(h)U_{I}(g);suppf,h\in W^{\prime }\right\} 
\nonumber
\end{eqnarray}
i.e. the two wedge-localized algebras (W denotes the right wedge) are
generated by free fields ``twisted'' by global $U(1)$ symmetry
transformation of angle $g$ (coupling constant)\footnote{%
The equality of the ${\cal A}(W)$ net (\ref{alg}) to the net obtained by the
subsequent modular method adapted to the Federbush model is not a very easy
matter.}. This follows from the observation that if the x is restricted to W
one may replace the exponential in $\psi _{I}$ (which represents a left half
space rotation) by the full rotation since the exponential of the right
halfspace charge is already contained in the right free fermion algebra etc.
The following unitarily equivent description of the pair $A(W),A(W^{\prime
}) $ has a more symmetric appearance under the parity symmetry $\psi
_{I}(t,x)\leftrightarrow \psi _{II}(t,-x)$:\vspace{0in} 
\begin{eqnarray}
{\cal A}(W) &=&alg\left\{ \psi _{I}^{(0)}(f)U_{II}(\frac{g}{2}),\psi
_{II}^{(0)}(h)U_{I}(-\frac{g}{2});suppf,h\in W\right\} \\
{\cal A}(W^{\prime }) &=&alg\left\{ \psi _{I}^{(0)}(h)U_{II}(\frac{g}{2}%
),\psi _{II}^{(0)}(f)U_{I}(-\frac{g}{2});suppf,h\in W^{\prime }\right\} 
\nonumber
\end{eqnarray}
The computation \cite{SW} of the scattering matrix $S_{s}$ from (\ref{local}%
) is most conveniently done by Haag-Ruelle scattering theory \cite{Haag}: 
\begin{eqnarray}
S_{s}\left| \theta _{1}^{I},\theta _{2}^{II}\right\rangle
&=&S_{s}^{(2)}\left| \theta _{1}^{I},\theta _{2}^{II}\right\rangle =e^{i\pi
g}\left| \theta _{1}^{I},\theta _{2}^{II}\right\rangle \\
S_{s}^{(n)} &=&\prod_{pairings}S_{s}^{(2)}  \nonumber
\end{eqnarray}
These formulae (including antiparticles) can be collected into an operator
expression \cite{SW} : 
\begin{equation}
S_{s}=\exp i\pi g\int \rho _{I}(\theta _{1})\rho _{II}(\theta
_{2})\varepsilon (\theta _{1}-\theta _{2})d\theta _{1}d\theta _{2}
\end{equation}
Where $\rho _{I,II}$ are the momentum space charge densities in the rapidity
parametrization.

The surprising simplicity of the wedge algebra as compare to say double cone
algebras consists in the fact that one can choose on-shell generators. This
is not just a consequence of the on-shell particle number conservation, but
requires the energy independence of the elastic scattering. Another example
is the Ising field theory.

For the more interesting factorizing models with energy dependent S-matrices
we find the distinction between diagonal and nondiagonal $S_{s}$very
helpful. Examples for the former are the $Z_{N}$-models whereas the various
Gross-Neveu models have nondiagonal two-particle S-matrices.

Considering first the diagonal situation $S^{(2)}=e^{i\delta (\theta )}$ $($%
with $\delta $ denoting the scattering phase shift), we define (assuming for
simplicity charge neutrality) operators $b$ in terms of the incoming $a^{\#}$%
: 
\begin{eqnarray}
b(\theta ) &:&=a(\theta )\exp -i\pi \int_{-\infty }^{\theta }\delta (\theta
-\theta ^{\prime })a^{*}(\theta ^{\prime })a(\theta ^{\prime })d\theta
^{\prime }:  \label{b} \\
b^{J}(\theta ) &:&=a^{*}(\theta )\exp -i\pi \int_{\theta }^{+\infty }\delta
(\theta -\theta ^{\prime })a^{*}(\theta ^{\prime })a(\theta ^{\prime
})d\theta ^{\prime }  \nonumber
\end{eqnarray}
It is easy to check that they fulfill the relation of the Zamolodchikov
algebra \cite{Zam}: 
\begin{equation}
b(\theta )b(\theta ^{\prime })=S^{(2)}(\theta -\theta ^{\prime })b(\theta
^{\prime })b(\theta )\,\,\,\,etc.
\end{equation}
In fact they define a representation in the physical in-Fock space. As we
will see below the main issue is not the algebra itself, but rather the
construction of an in-representation in case where $S^{(2)}$ is nondiagonal
and an analogue formula to (\ref{b}) does not seem to be available.
Combining the $b$-operator with its ``j-adjoint'' $b^{J\text{ }}$, we obtain
the following nonlocal but TCP-invariant (and therefore weakly local)
operator: 
\begin{equation}
B(x)=\int (b(\theta )e^{-ipx}+b^{J}(\theta )e^{ipx})d\theta
\end{equation}
where $b^{J}$ has been defined in (\ref{b}). Its use facilitates greatly the
construction of modular wedge localized states because it turns out that the
closure of the real space: 
\begin{equation}
\int f_{n}(x_{1},....x_{n}):B(x_{1})....B(x_{n}):\Omega
,\,\,\,\,suppf_{n}\in W^{\otimes n},\,\,f_{n}\,\,real
\end{equation}
is the desired ${\cal H}_{R}^{(n)}(W)$ i.e. solves the -1 eigenvalue
equation for the ``interacting'' Tomita operator $S=S_{s}S_{0}.$ The
formfactors of local fields: 
\begin{equation}
^{out}\left\langle p_{1}^{\prime },...p_{m}^{\prime }\left| A(0)\right|
p_{1},...p_{n}\right\rangle ^{in}
\end{equation}
can be expressed in terms of a vacuum expectation value which involves n+m
fields $B^{\#}$ and one (at this instance still unknown) local field $A:$%
\begin{equation}
\left\langle
B(x_{1})^{\#}...B(x_{m})^{\#}A(x)B(x_{m+1})^{\#}...B(x_{n+m})^{\#}\right%
\rangle
\end{equation}
Introducing the natural parametrization for the 2-dim wedge W in terms of
the radius r and the x-space rapidity $\chi ,$ the KMS property reads: 
\begin{eqnarray}
&&\left\langle B(r_{1},\chi _{1})^{\#}...B(r_{m},\chi
_{m})^{\#}A(x)B(r_{m+1},\chi _{m+1})^{\#}...B(r_{n+m},\chi
_{n+m})^{\#}\right\rangle  \label{KMS} \\
&=&\left\langle B(r_{2},\chi _{2})^{\#}...B(r_{m},\chi
_{m})^{\#}A(x)B(r_{m+1},\chi _{m+1})^{\#}...B(r_{1},\chi _{1}-2\pi
i)^{\#}\right\rangle  \nonumber
\end{eqnarray}
for $A=1$ and $n+m=4$ we obtain: 
\begin{eqnarray}
&&\left\langle B(r_{1},\chi _{1}+i\pi )B(r_{2},\chi _{2})B(r_{3},\chi
_{3})^{\#}B(r_{4},\chi _{4})^{\#}\right\rangle \\
&=&\left\langle B(r_{2},\chi _{2})B(r_{3},\chi _{3})^{\#}B(r_{4},\chi
_{4})^{\#}B(r_{1},\chi _{1}-i\pi )\right\rangle  \nonumber
\end{eqnarray}
Expressing this in terms of the momentum space rapidity variables we get a
special case of the well-known crossing symmetry for the S-matrix: 
\begin{equation}
S_{s}(\theta )=S_{s}(i\pi -\theta )
\end{equation}
The more general case of particles$\neq $antiparticles relates $S_{s}^{pp}$
with $S_{s}^{p\overline{p}}$ and can be similarly discussed. A
generalization to higher dimensions for which the analogon of the auxiliary
fields $B$ is unknown (and for which the presence of infinitely many
L-transformed wedges gives rise to a consistency problem) is presently
unknown, although the validity of the KMS property with its strip
analyticity and the (not so well established) on shell crossing symmetry are
certainly not independent analytic ``symmetries'' of QFT. They are both
fundamentally linked with the issue of antiparticles. I expect that progress
on this subject will relate modular localization to the ill-understood on
shell analytic properties of QFT.

The most interesting case is the nondiagonal case where the ``b-trick'' does
not seem to be available. In that case our strategy will be to abstract a
complete set of analytical properties for the wedge-localized n-particle
wave functions: 
\begin{eqnarray}
f^{A}(\theta _{1},...\theta _{n})\,\,\, &=&\left\langle \Omega \left|
A\right| p_{1},...p_{n}\right\rangle ^{in}=\left\langle \Omega \left|
A\right| a^{*}(p_{1}),...a^{*}(p_{n})\right\rangle \,\,\,\,\,\,for\,\,\theta
_{1}>...>\theta _{n} \\
&=&\left\langle \Omega \left| A\right| b^{*}(p_{1}),...,b^{*}(p_{n})\Omega
\right\rangle \,\,\,\,\,\,\,\,  \nonumber
\end{eqnarray}
where the ordering has been chosen because such incoming particles do not
cross in the future, i.e. $S_{s}$ acts trivially and the purpose of the
second line is only to suggest the right boundary conditions for the
meromorphic function in arbitrary $\theta $-order: 
\begin{eqnarray}
\,\,\,\,\,f^{A}(\theta _{P(1)},...,\theta _{P(n)}) &=&\lim_{\func{Im}%
z_{P(1)}>...>\func{Im}z_{P(n)}\rightarrow 0}f^{A}(z_{1},...,z_{n})
\label{c.r.} \\
f^{A}(...,\theta _{i},\theta _{j},...) &=&f^{A}(...,\theta _{j},\theta
_{i},...)S_{s}^{(2)}(\theta _{i}-\theta _{j})  \nonumber
\end{eqnarray}
The second important suggestive role of the b-representation stems from the
(right wedge) KMS-property (\ref{KMS}): 
\begin{eqnarray}
&&\left\langle \Omega \left| AB^{*}(r,\chi _{1})...B^{*}(r_{n-1},\chi
_{n-1})B^{*}(r_{n},\chi _{n}-2\pi i)\right| \Omega \right\rangle \\
&=&\left\langle \Omega \left| B^{*}(r_{n},\chi _{n})AB^{*}(r,\chi
_{1})...B^{*}(r_{n-1},\chi _{n-1})\right| \Omega \right\rangle
,\,\,\,\,\,\,A\in {\cal A}(W)  \nonumber
\end{eqnarray}
or equivalently: 
\begin{eqnarray}
&&\left\langle \Omega \left| AB^{*}(r,\chi _{1})...B^{*}(r_{n-1},\chi
_{n-1})B^{*}(r_{n},\chi _{n}-\pi i)\right| \Omega \right\rangle \\
&=&\left\langle \Omega \left| B^{*}(r_{n},\chi _{n}+i\pi )AB^{*}(r,\chi
_{1})...B^{*}(r_{n-1},\chi _{n-1})\right| \Omega \right\rangle  \nonumber
\end{eqnarray}
The {\it on-shell nature of the B-fields} permits an easy transformation of
this KMS condition to momentum rapidity space (p$_{i}\rightarrow \theta _{i}$%
): 
\begin{eqnarray*}
&&\left\langle \Omega \left| Ab^{*}(\theta _{1})....b^{*}(\theta
_{n-1})b^{*}(\theta _{n}+i\pi )\right| \Omega \right\rangle \\
&=&\left\langle \Omega \left| b^{J}(\theta _{n}-i\pi )^{*}Ab^{*}(\theta
_{1})....b^{*}(\theta _{n-1})b^{*}(\theta _{n}+i\pi )\right| \Omega
\right\rangle
\end{eqnarray*}
The quantum field theoretical interpretation of the fouriertransform of the
on shell KMS condition is: the analytic continuation to the upper rim of the
strip of the wedge-localized wave function of $A$ is the crossed matrix
element: 
\begin{eqnarray}
&&\left\langle \Omega \left| b^{J}(\theta _{n}-i\pi )^{*}Ab^{*}(\theta
_{1})....b^{*}(\theta _{n-1})\right| \Omega \right\rangle \\
&=&\left\langle \Omega \left| b(\theta _{n})Ab^{*}(\theta
_{1})....b^{*}(\theta _{n-1})b^{*}(\theta _{n}+i\pi )\right| \Omega
\right\rangle  \nonumber
\end{eqnarray}
By iteration we obtain: 
\begin{eqnarray}
&&\left\langle \Omega \left| Ab^{*}(\theta _{1})....b^{*}(\theta
_{k-1})b^{*}(\theta _{k}+i\pi )...b^{*}(\theta _{n}+i\pi )\right| \Omega
\right\rangle \\
&=&\left\langle \Omega \left| b(\theta _{k})...b(\theta _{n})Ab^{*}(\theta
_{1})....b^{*}(\theta _{k-1})\right| \Omega \right\rangle  \nonumber
\end{eqnarray}
For nonselfconjugate particles the crossed operators on the left must be
replaced by antiparticle annihilation operators. If the $S$-matrix has
matrix indices (particle multiplets with nondiagonal 2-particle S-matrix),
the $f^{A}$ carry n matrix indices. Denoting by $\theta $ now the rapidity
together with the multiplet index and using $\bar{\theta}$ for the rapidity
of the antiparticle we have: 
\begin{eqnarray}
f^{A}(\bar{\theta}_{n},\theta _{1},\theta _{2},...,\theta _{n-1})
&=&\left\langle \Omega \left| b^{J}(\theta _{n}-i\pi )^{*}Ab^{*}(\theta
_{1})....b^{*}(\theta _{n-1})\right| \Omega \right\rangle \\
&=&f^{A}(\theta _{1}+i\pi ,\theta _{2},...,\theta _{n})=f^{A}(\theta
_{n}-i\pi ,\theta _{2},...\theta _{1})  \nonumber
\end{eqnarray}
In terms of the original functions $f^{A}(\theta _{1},...\theta _{n})$ it is
very instructive to group the boundary relations as follows: 
\begin{equation}
\overline{f^{A}(\theta _{1}+i\pi ,...,\theta _{n}+i\pi )}=f^{A}(\bar{\theta}%
_{1},...,\bar{\theta}_{n})  \tag{ML}
\end{equation}
\begin{equation}
f^{A}(\theta _{n}-2\pi i,\theta _{1},...\theta _{n-1})=f^{A}(\theta
_{1},\theta _{2},...\theta _{n})  \tag{KMS}
\end{equation}
\begin{equation}
f^{A}(\bar{\theta}_{n},\theta _{1},...,\theta _{n-1}):=f^{A}(\theta
_{1},\theta _{2},...\theta _{n}+i\pi )  \tag{Def}
\end{equation}
The last equation is the (iterative) definition of the crossed channel of
the (right most) particle and the second relation is the analytic property
of crossing, alias KMS. The first equation is the most general of all: the
spatial modular localization relation. Although states of the form $A\Omega $
with $A\in {\cal A}_{s.a.}(W)$ fulfill this relation, it also holds for the
rapidity wave functions of any vector $\psi $ which lies in the Tomita graph
closure (i.e. in the natural wedge localized space) of ${\cal A}(W)\Omega $.
In that case the relation reads: 
\begin{equation}
\overline{\left\langle \psi \mid \theta _{1}+i\pi ,...\theta _{n}+i\pi
\right\rangle ^{out}}=\left\langle \psi \mid \theta _{1},...\theta
_{n}\right\rangle ^{in}
\end{equation}
Unlike the special case of $\psi =A\Omega $ there are no furthergoing
analytic properties beyond the $\pi $-strip analyticity (i.e. no globally
meromorphic functions.

The other relation involving KMS property are only meaningful on modular
localized vectors of the form $A\Omega .$ The additional relation for
factorizable systems which only hold in the absence of real particle
relation are those which express the n! different boundary values
distinguished by the order of $\theta ^{\prime }s$ i.e. their deviation from
the reference order $\theta _{1}>...>\theta _{n}$ via a commutation relation
(\ref{c.r.}). The above relations ML, KMS together with the definition Def
and the commutation relations (\ref{c.r.}) appear precisely in this form as
postulates in the recent work of Babujian Fring and Karowski \cite{BFK}\cite
{S2}. Since the values on the upper strip boundaries are related by known
analytic functions which depend on the ordering of the $\theta ^{\prime }s$
to the corresponding lower boundaries one may say that modular theory
explains why many of the analytic techniques for factorizing models in
momentum space are very similar to x-space analytic properties in Wightman's
formulation of QFT in particular of conformal QFT.

There is one more equation in the BFK scheme of axioms which from or point
of view is a kind of one particle structure property relating the wave
functions or formfactors for different n \cite{BFK}\cite{S2}: 
\begin{equation}
f^{A}(\bar{\theta}_{n},\theta _{1}...,\theta _{n-1})\stackunder{\theta
_{1}\rightarrow \theta _{2}+i\pi }{\simeq }\frac{2i}{\theta _{n}-\theta
_{n-1}-i\pi }f^{A}(\theta _{1},...,\theta _{n-2})(1-S_{n-2,n-1}...S_{1k,n-1})
\label{cl}
\end{equation}
These ``kinematical poles'' outside the physical strip (but approaching its
boundary have their physical origin in the so called one particle structure
in QFT as first noticed by Stueckelnberg. In factorizing theories they take
their above specific form. For nonselfconjugate particle situations the
residuum on the right hand side is only different from zero if permitted by
the charge superselection rules (i.e. one of the right hand particles must
be an particle which can be contracted with the antiparticle symbolized by $%
\bar{\theta}_{n}$). Inside the physical region the $S$-matrix may have bound
state poles. In that case one has to enlarge the scattering space by the
Fock space of these new incoming particles. Their presence is also felt in
the formfactors of the original particles; they have bound state poles and
their residua are determined by the S-matrix poles and crossing (KMS). We do
not need the concrete formulae from the one particle structures since we
will not enter concrete model constructions in this paper. The method of
finding the wedge localized Hilbert spaces is now the following. In case of
particle multiplets (say SU(n)) we first decompose the vector $A\Omega $
into irreducible representations (highest weight vectors). Assuming that we
already have a special solution $f_{w}^{A}$ for the highest weight w, the
most general solution which defines the localized wave function spaces is: 
\begin{equation}
f_{w}^{A}:{\cal H}_{0}^{(n)}\rightarrow {\cal H}_{w}^{(n)}
\end{equation}
\begin{eqnarray}
g_{w}(\theta _{1},...,\theta _{n}) &=&f(\theta _{1},...\theta
_{n})f_{w}^{A}(\theta _{1},...\theta _{n})  \label{comp} \\
f &\in &{\cal H}_{0}^{(n)},\,\,\,g_{w}\in {\cal H}_{w}^{(n)}  \nonumber
\end{eqnarray}
Here $f\in {\cal H}_{0}^{(n)}$ ranges through the space of solutions of the
above equations with trivial scattering S-matrix. It is easy to see that
this space consists of periodic functions with no poles but an arbitrary
number of zeros. The space is identical to that of formfactors of local
operators in free theories. There are now two crucial questions to be
answered:

\begin{enumerate}
\item  Are the wedge algebras ${\cal A}_{1}(W)$ and ${\cal A}_{2}(W)$ of two
theories the same if the spaces (i.e. their $\theta $-space wave functions)
are identical ${\cal A}_{1}(W)\Omega ={\cal A}_{2}(W)\Omega $? Since both
wedge algebras are hyperfinite type III$_{1}$ factors, they are not only
algebraically isomorphic but also unitary equivalent: ${\cal A}_{2}(W)=U%
{\cal A}_{1}(W)U^{*}.$ So the question is $U\in {\cal A}^{\prime }(W)$?

\item  How can one obtain a special solution of the full equations for
factorizing models?
\end{enumerate}

The first question has a surprisingly simple answer. If the algebras on the
vacuum are identical, then (thanks to the crossing relations!) the action on
arbitrary incoming particle vectors is the same implying identity of the
algebras. This answers the question of uniqueness of the field theory for a
given admissable S-matrix i.e. the {\it uniqueness of the inverse problem}.
In particular it shows that $S=1$ with Bose- or Fermi-statistics has only
the free field solution. This was a long standing problem; only for zero
mass a solution was known \cite{BuF}.

The second question i.e. the existence for factorizable QFT is more
difficult. In the diagonal case it would be very easy to write down special
solutions, if it would not be for the pole structure. This is a problem for
which an efficient method still has to be found. It turns out that the more
difficult case of nondiagonal S-matrices for particle multiplets can be
reduced to the previous case by the {\it nested Bethe-Ansatz method}.
Depending on the weight, the solution can be shown to admit the algebraic
Bethe\footnote{%
We use the terminology {\it Bethe Ansatz} (or representation) whenever state
vectors of an interacting system are written as a superposition of products
of known (matrix) operators. In the case at hand these operators are the
collection (labeled by $\beta _{i}$) of \underline{$\theta $}-dependent
B-matrices which depend on the additional parameter $u.$ Whereas the
perturbative Feynman ``machine'' exists in a fully developed form, we
presently only have a rudimetary knowledge of the Bethe machine.}
representation \cite{BKZ}: 
\begin{eqnarray}
f_{1...n}(\underline{\theta }) &=&\sum_{\underline{u}}B_{1...n,\beta _{m}}(%
\underline{\theta },u_{m})...B_{1...n,\beta _{1}}(\underline{\theta }%
,u_{1})\Omega _{1...n}g^{\beta _{1}...\beta _{n}}(\underline{\theta },%
\underline{u}) \\
\Omega _{1...n} &=&e_{1}\otimes ...\otimes e_{1}\in V\otimes ...\otimes V
\end{eqnarray}
Let us restrict to SU(2) multiplets. In that case $V$ is 2-dimensional. The $%
B^{\prime }s$ are known matrix-valued functions acting on $\Omega _{1...n}$
which have an extra index $\beta .$ They are expressed in terms of a
suitably defined n-fold tensor product of the two-particle S-matrix \cite
{BKZ} i.e. in terms of the so called algebraic Bethe Ansatz. The number m of
B-factors is related by a simple formula to the highest weight of the
representation. If we chose the coefficients g as: 
\begin{equation}
g^{\beta _{1}...\beta _{m}}(\underline{\theta },\underline{u}%
)=\prod_{i=1}^{n}\prod_{j=1}^{m}\psi (\theta _{i}-u_{j})\prod_{1\leq i<j\leq
m}\tau (u_{i}-u_{j})f^{(1)(\beta _{1},...\beta _{m})}
\end{equation}
with $\psi $ and $\tau $ being ratios of $\Gamma $-functions and $%
f^{(1)(\beta _{1}...\beta _{m})}(\underline{u})=$constant \cite{BKZ}, then $%
f_{1...n}(\underline{\theta })$ solves the Riemann-Hilbert problem i.e. the
set of boundary conditions without the formulae which relate the residuum of
one-particle poles to lower particle $f^{\prime }s\footnote{%
For n-particle multiplets obeying the SU(n) symmetry, the $f^{(1)(\beta
_{1},...\beta _{m})}$ for fixed $\beta ^{\prime }s$ are matrices acting on a
tensor space $V^{(1)}\otimes ...\otimes V^{(1)}$ where $dimV^{(1)}=n-1.$ In
that case one needs a nested Bethe representation, i.e. one has to repeat
the Bethe Ansatz n-1 times.}.$ The problem is not to find a special solution
with a given pole structure, but rather the tuning of to the residua as
demanded by the formfactor interpretation. This problem is absent in the
case of trivial S-matrix $S=1$. We expect that this tuning is not necessary
if we are only interested in the position of the wedge localizes Hilbert
space ${\cal H}(W)$ inside the Fock space.

\begin{conjecture}
The dense set of wedge localized states is the direct sum of n-particle
components (\ref{comp}) where $f^{A}$ now denotes a special solution of the
multi-strip n-variable Riemann-Hilbert problem without the residuum
condition on the pole structure.
\end{conjecture}

Our physical picture underlying this conjecture is that the S-matrix $S_{s}$
and hence the wedge Tomita operator $S(W)$ commutes with the (on shell)
particle number. But for interacting theories this cannot hold for smaller
regions i.e. $\left[ S(O),{\bf N}\right] \neq 0$ for double cones. This can
be verified in the Federbush model and corresponds to the physical picture
that localization beyond the wedge region requires ``virtual particle
creation''. This is also the reason why factorizable models have such a rich
virtual particle structure\footnote{%
For example their two-point function always have momentum space
contributions above the Wigner one particle mass shell.} even though there
is no on shell creation. In fact the above conjecture suggests to sharpen
the picture of the local Wightman domain (or algebraic range ${\cal A}%
(W)\Omega $) as: 
\begin{eqnarray}
{\cal A}(W)\Omega &=&{\cal D}(W)\subset {\cal H}(W) \\
{\cal D}(W) &=&\bigcup_{W\supset O_{1}}domS(O_{1})  \nonumber
\end{eqnarray}
In this conjectured formula ${\cal D}(W)$ denotes the (open) union of all
domains of Tomita operators for double cones ranging over all unit double
cones inside W. ${\cal H}(W)$ is the (closed in the temperature topology of
section 3) Hilbert space of modular wedge localization. Accepting the
correctness of this conjecture, the relation between the n-particle
components originates from the noncommutativity of the Tomita operators $%
S(O) $ with the particle number and ${\cal D}(W)$ inherits this property.
This is the origin of the rich virtual particle structure of local fields
and of the interrelation of the n-particle components for different n of
vectors obtained by applying these fields to the vacuum.

In case of existence of the Zamolodchikov operators $B$ the component
structure of ${\cal H}(W)$ is obvious since the commutation structure of the
creation and annihilation parts of $B$ with the particle number operator is
the same as that of a free field and ${\cal H}^{(n)}(W)$ is obtained by
n-fold application of the smeared $B^{\prime }s$ where the support of the
smearing function is in W\footnote{%
The $B^{\prime }s$ can however not be used in order to generate Hilbert
spaces for smaller than wedge localizations. Their auxilary role is strictly
confined to modular localized wedge spaces.}.

The most interesting part is passing from the localized states to {\it local}
fields and algebras. Let me only explain this for the (diagonal) case where
the $B^{\prime }s$ have been explicitly constructed. These fulfill all the
modular covariance properties with respect to $J,\Delta ^{i\tau }$ and $S$
except that the vacuum is not separating for the B-algebra (related to their
nonlocality). Consider now operators of the following form: 
\begin{eqnarray}
A &=&\sum \int g^{(\#)}(y_{1},...y_{n}):B^{\#}(y_{1})...B^{\#}(y_{n}): \\
\,\,\,\,\,W &\ni &\,\,supp\,g^{(\#)}(x;y_{1},...y_{n})  \nonumber
\end{eqnarray}
Here we use \# to denote the operator or its Hermitian adjoint and the
superscript on g indicates that the coefficient functions depend on which $%
B^{\prime }s$ are replaced by Hermitian conjugate. Note that, contrary to
free fields, the $B^{\prime }s$ are not commutative under Wick-ordering. As
a result of the on shell nature of the $B^{\prime }s,$ the above formula
gives upon Fourier transformation: 
\begin{equation}
A=\sum \int \tilde{g}_{n}^{(\#)}(p_{1},....p_{n}):\tilde{B}^{\#}(p_{1})....%
\tilde{B}^{\#}(p_{n}):d\theta _{1}...d\theta _{n}
\end{equation}
Where now \# stands for the momentum space annihilation or creation part.
The $J$ (TCP) -covariance and the $\Delta ^{it}$ covariance gives a simple
restriction on the coefficient functions. The only property of $A$ which is
missing in order to qualify as a generator for a wedge algebra in standard
position is separability. This can however be enforced if the pure
annihilation part of the n$^{th}$ contribution $\tilde{g}$ is related to the
creation part so that a vanishing of the creation parts without a vanishing
of the annihilation contribution is impossible, hence an $A$ which
annihilates $\Omega $ vanishes as an operator. Precisely this is guarantied
if the different frequency parts of the n$^{th}$ contribution are identified
with the various would be n-particle formfactors of $A,$ as required by the
definition of the $B^{\prime }s.$ As already stated, separability of the
wedge $A(W)$-algebra together with cyclicity and the above covariance
property will enforce the ``standardness'' of its representation relative to
the vacuum. Since $S$ is then the Tomita operator of this separable
subalgebra of the algebra generated by the $B^{\prime }s$ restricted to the
wedge, the relative locality of the algebra for the opposite wedge is a
consequence of modular theory. This separability argument is a significant
analytic simplification and conceptual clarification as compared to the
existing recipe for checking locality \cite{Smi}. The previously presented
Federbush model offers a nice explicit illustration for the usefulness of
the $B^{^{\prime }}s$ in the construction of the local wedge algebras and
the local fields $\psi ^{I,II}.$

Here the question remains if modular wedge localized states in factorizable
models can always be represented in terms of nonlocal on shell operators $B$
with positive and negative frequency parts which fulfill a Zamolodchikov
algebra with a nondiagonal $S^{(2)}$-structure constants. I conjecture that
this is the case. The fact that the $\Psi _{n}$ for different n are related
by (\ref{cl}) is favorable for such a conjecture, but even if one can show
existence there is still the problem of finding the explicit representation
in terms of incoming fields. It would be highly desirable to have a
reconstruction theorem for the $B$-operators from the existence of wedge
localized n-particle vectors and their formfactors. Closely related to the
existence of the $B^{\prime }s$ is the existence of a ``modular M\o ller
operator'' U defined by: 
\begin{equation}
UJ_{0}U^{-1}=S_{s}J_{0}=J  \label{U}
\end{equation}

The existence of this ``square root'' of the S-matrix $S_{s}$ is easily
shown to be equivalent to the existence of an ``interacting'' wedge algebra: 
\begin{equation}
{\cal A}(W)=U{\cal A}_{in}(W)U^{-1}
\end{equation}
which together with the vacuum vector belongs to the modular object $%
J,\Delta ^{i\tau }.$ In applying this modular M\o ller operator to the free
field coordinates of incoming fields it is very important to realize that
the point $x$ of the resulting U-transformed field does nor denote
localization around $x$. Although this field transforms in the standard
manner under Poincar\'{e} transformations, its localization is spread (the
``fuzziness'' in section 3) all over the wedge. Compact localization can
only be achieved by intersecting wedge algebras, and not by further
physically interpretable covariant transformations. Furthermore the $U$
(against naive expectations) cannot commute with the Poincar\'{e}
transformations since otherwise the ${\cal A}(W)$ net is the same (i.e. the
local algebras are unitarily equivalent with the same unitary) to the ${\cal %
A}_{0}(W)$ net \footnote{%
In fact, as was shown by Wollenberg \cite{Wo}, the existence of such a $U$
is equivalent to (${\cal A}(W),U^{*}\Omega )$ having the modular object ($%
J,U^{*}\Delta ^{\frac{1}{2}}U$).}. However the requirement that the
commutant be geometric i.e. $JA(W)J=PA(W)P^{-1},$ where $P$ is any geometric
transformation linking the right with the left wedge (in d=1+1 it is a
parity symmetry in higher d there are also $\pi $-rotations). We will
investigate the uniqueness (modulo the above freedom) and the physical
significance of this modular M\o ller operator $U,$ as well as its
computability in factorizable models and its possible relation to the
modular localization equations and the in-representation of the
Zamolodchikov algebra structure in a separate paper \cite{SWi}.

\section{ General Interactions and Modular Localization.}

We mentioned already that e.g. ``free'' d=1+2 anyons do not have a Fock
space structure. So the functorial method of section 2 relating the Wigner
space directly to operator algebras will not work, even though there is no
genuine interaction. Since the inner products of multiparticle momentum
states and the action of the Poincar\'{e} group is however explicitely
known, one can start a program of computing n-particle wedge localization
spaces and think about spaces of spacelike cone localization spaces and the
associated operator algebras. This will not be done here, the main reason
being that presently there are no substantial results on this program.

Instead we will address the important question whether the concept of
modular localization can be expected to lead to a nonperturbative approach
for d=1+3 interacting theories of Fermions and Bosons. Let us try to think
about a scenario consisting of three steps.

\begin{enumerate}
\item  Start with an auxiliary $S^{(0)}$-matrix which is ${\cal P}$%
-invariant (including $TCP$ and $J$ reflections) and fulfills some weak
cluster properties. An example would be: 
\begin{equation}
S^{(0)}=e^{i\eta },\,\,\,\,\eta =g\int :A^{(0)}(x)^{4}:d^{4}x
\end{equation}
where $A^{(0)}$ is a scalar free field. We use this initial S-matrix for
baptizing the theory ``$A^{4}$-theory''

\item  Try to construct a modular localization space ${\cal H}_{R}(W)$: 
\begin{equation}
S^{(0)}J\Delta ^{\frac{1}{2}}\psi =\psi ,\,\,\,\,\psi \in H_{Fock}^{in}
\end{equation}
without on shell conservation laws (which before lead to the Bethe Ansatz
structure), this seems to be a tough mathematical problem.

\item  Try to obtain more refined localization spaces. If multilocal spaces
spaces could be constructed then the ideas of scattering theory could lead
to an $S^{(1)}$-operator which would agree with the starting S-matrix $%
S^{(0)}$ if the the theory would be local (which it is not). Use $S^{(1)}$
as a new modular input. By playing the modular S-operator iteratively
against a scattering operator one hopes that the iteration $%
S^{(0)},S^{(1)}.....$could lead to a limiting local situation for which the
modular and the scattering $S$ agree.
\end{enumerate}

The difference of such a hypothetical iteration to a deformation approach as
standard perturbation theory would be that in this approach the operator
properties (except locality) would be valid in every step. We expect the
modular M\o ller operator U (\ref{U}) to play an important role, since its
existence is equivalent to the existence of an interacting wedge algebra.

As expected from the infrared aspects of standard perturbation theory, the
zero mass theories also show a special behavior in the modular localization
approach. In the latter case one has to face the physically more serious
problem of a missing free reference theory resulting from the vanishing LSZ
asymptotic limits of ``infraparticles''\footnote{%
``Infraparticles'' are ``below'' Wigner particles in the sense that their
contribution to the two-point function is less singular than the Wigner (on
shell) contribution.}. This problem already arises in chiral conformal QFT
with noncanonical scaling dimensions. The dialectic tension between analytic
simplicity and conceptual complexity of massless QFT is one of the
fascinating phenomena of contemporary research in QFT.

\section{Resume and Outlook}

Low-dimensional models of QFT, as e.g. chiral conformal theories or massive $%
d=1+1$ models, have, apart from possibly condensed matter applications, no
direct use in physics. However they are excellent laboratories for
theoretical ideas about elementary particle physics. The standard
perturbative approach has not only led to the well-known successes, but also
created some folklore about nonperturbative aspects which are sometimes not
entirely correct and even prejudicial. In the following we compile few of 
{\bf these incorrect statements} which low-dimensional models solved by
nonperturbative methods are able to correct:

\begin{itemize}
\item  ``The existence of QFT is endangered by bad short-distance behavior''.
\end{itemize}

$D=1+1$ soluble models show that this is not so. This is already clear for
the simplest of them, as the Federbush model used in these notes, for which
one does not even have to invoke the modular localization and the related
bootstrap-formfactor-program. By changing a deformation parameter (the
coupling constant), the inverse short distance powers may be chosen as large
as desired. In the algebraic approach, based on nets of local algebras, the
short distance properties are hidden and only appear via associated scaling
algebras. It is presently not entirely clear which short-distance aspects
are intrinsic net features, and which are attributes of particular
generating ``field coordinates''.

\begin{itemize}
\item  ``QFT suffers from short distance divergencies''.
\end{itemize}

This is closely related to the previous folklore. Is not even correct in
perturbation theory. The divergencies in Feynman integrals are a result of
the Lagrangian (or canonical) quantization methods which starts from a
slightly illegitimate picture about the nature of quantum fields and which
needs repair by infinite renormalization. As Poincar\'{e} and Lorentz showed
at the beginning of the century, these infinities are genuine in the
classical theory because the particle picture has to be imposed on the
classical field theory. However in local quantum physics, the particle
properties are part of the Poincar\'{e} transformation properties of fields
and as such follows from those. Indeed, if one changes the formulation
slightly and aims first at a formally unitary localization function
dependent perturbatively accessible interaction operator $S(g)$ in an
auxiliary Fockspace which fulfills the Bogoliubov axiomatic, then the
problem of infinities is traded with the problem of Hahn-Banach extension of
time ordered functions which are originally defined for noncoinciding
arguments. An even more convincing finite methods is the split-point
treatment of the nonlinear terms in operator field equation. Quantum fields
are singular for physical reasons and to demand (naive) finiteness in a
Lagrangian or functional integral formulation or any other quantization
(parallelism to classical physics) is self-defeating. Some of these
manifestly finite methods are not very practical and therefore I recommend
to stay with the original Feynman methods but to avoid to draw wrong
philosophical conclusions from it. The nonperturbative modular localization
method is obviously free of divergencies and even in the higher dimensional
cases without an initial physical S-matrix where one expects at best an
iterative procedure to succeed, one does not see how infinities could
possibly enter. Certainly there is no place for short-distance infinities in
a net approach, it is too far removed from quantization ideas which always
tend to sneak in the classical relativistic particle problems of
Poincar\'{e} and Lorentz and require the repair called (infinite)
renormalization. Neither is there a place for time-ordered products or a
Dirac type ``interaction picture''. Instead one meets a new structure:
auxiliary ``on shell'' fields and related algebras which are between the
local interacting and the free incoming fields and which in special cases
define a physical realization of the Zamolodchikov algebra and in more
general cases seem to be related to a Bethe-Ansatz structure.

\begin{itemize}
\item  ``Lagrangians and actions are indispensable tools of constructive
QFT''.
\end{itemize}

With the exception of the Federbuch model and some similar almost trivial
models, none of the models constructed by the modular bootstrap-formfactor
method has been constructed by using such properties. In fact, even if one
can affiliate a Lagrangian with such a model, as in the case of the
Sine-Gordon or Thirring Lagrangian, it is mainly used for ``baptizing'' the
model in a conventional manner and plays no role in its nonperturbative
construction. In most cases, especially those without deformable coupling
constants, a Lagrangian is not even known.

\begin{itemize}
\item  ``Supersymmetry is a rich symmetry with many physical consequences''
\end{itemize}

Whereas in low dimensions the issue of internal/external symmetries become
inexorably intermingled, in the d=1+3 world there seemed to be no nontrivial
way to marry internal and space-time symmetries. In some way supersymmetry
seemed to lead to such mildly nontrivial looking marriage. Supersymmetry
turned out to be a mathematically apparently powerful symmetry whose
physical value is however highly questionable, to say the least. Using our
best classification methods for symmetry\footnote{%
The space-time symmetry can be directly constructed from the modular groups
of the family of wedge algebras, whereas the endomorphisms associated with
charged representations leading to internal symmetries, fulfill the Takesaki
``devissage'' property with respect to the Tomita-Takesaki modular structure.%
}, namely the Tomita-Takesaki modular, there is no trace of supersymmetry.
In order to see that we are dealing here with an accidental symmetry, i.e. a
symmetry of the field algebra which is not related to any additional
physical insight obtained beyond the mentioned method of classifying charges
we do not even have to invoke the modular bootstrap-formfactor program. A
glance at the simpler situation in chiral conformal field theory already
shows that e.g. in the case of the tricritical Ising model and similar
supersymmetric models, the supersymmetric formalism does not play any role
neither in their definition nor in their solution as members of a discrete
or continuous family of non-supersymmetric models. They are also not in any
way distinguished by short distance or other properties from their
nonsupersymmetric neighbors.. Of accidental symmetries one expects of course
instabilities under perturbation. Indeed, it has been shown recently that
super-symmetry suffers a ``collapse'' in temperature KMS states, a situation
totally different from any internal or genuine external (example
Lorentz-symmetry) symmetry which suffers at most a spontaneous symmetry
breaking related to the formation of phases in the theory of phase
transitions. Supersymmetry has been used to allege the perturbative
existence of a $d=1+4$ conformally invariant gauge theory. Even within the
context of perturbation theory such nontrivial gauge invariant correlation
functions in $d=1+3$ would be very interesting if not to say sensational.
But within its more than 15 years of folkloric existence, nobody has
calculated (or been able to calculate) such correlation functions, even in
lowest order i.e. those calculations which have been done for standard
theories and (nonsupersymmetric) gauge theories. Here one has some reason to
suspect something underneath the carpet.

Our criticism concerning the alleged rich physical content of supersymmetry
in no way applies to its use in mathematics.

We now turn to some {\it new results and problems}

\begin{itemize}
\item  {\bf \ ``Cooking recipes'' of the bootstrap-formfactor program permit
now a more profound understanding.}
\end{itemize}

In a tour de force Smirnov has compiles a list of formal requirements which
are designed to allow the extension of the Karowski-Weisz work on
two-particle form factors to arbitrary many particles. He demonstrated the
viability of these axioms by computing high formfactors of several models.
These ``axioms'' were recently brought into a physically somewhat more
transparent form within the so-called LSZ framework of QFT. By realizing
that the postulated ``crossing symmetry'' property is reducible to the KMS
property of modular wedge localized rector states, one found a way in which
this important construction approach becomes incorporated into algebraic
QFT. The traditional method of aiming at formfactors of fields has become
somewhat cumbersome. From our experience with the Federbush model we hope
that a direct construction (i.e. avoiding ``field coordinates'') of the
wedge algebra may be simpler.

\begin{itemize}
\item  {\bf \ Understanding of a Bethe Ansatz-like nonperturbative approach
to QFT.}
\end{itemize}

Bethe presented a technique by which certain low-dimensional problems in
lattice and continuum QFT could be solved. Although it always looked like a
shiny part of potentially impressive nonperturbation ``machine'', the Bethe
Ansatz method never reached the same maturity and perfection as the
perturbative Feynman ``machine''. Recently several authors have shown the
usefulness of (appropriately adapted) Bethe Ansatz techniques in the
bootstrap-formfactor program \cite{BKZ}\cite{BFK}.

It appears that the constructive use of the modular localization method may
now change the hole picture since it leads to Bethe-Ansatz structures in the
most direct conceptual manner. I am using the terminology (generalized)
Bethe-Ansatz machine also for the yet unknown analytic formalism behind the
more general scenario of the previous section which I expect to emerge as a
new constructive formalism in (non factorizable) local QFT.

\begin{itemize}
\item  {\bf Emergence of Geometry from domains of operators and ranges of
algebras.}
\end{itemize}

One of the most surprising aspects of this new way of looking at local
quantum physics is the encoding of geometric data in the net of domains of
the modular $S$ operators and the ranges of ${\cal A}(O)\Omega $. The
initially nongeometric principles of local quantum physics are in this
subtle way transformable into geometric properties. It is only through the
detailed understanding of this connection that one can utilize differential
geometry in physically correct way. Arguments of consistency of (global)
geometrical structures with one or the other form of QFT formalism is not
enough (example: the global vacuum structure for vacua which do not arise
from spontaneous symmetry breaking as in Seiberg-Witten duality discussion).

Finally we comment on possible {\bf relations of modular localization to
other approaches }as Alain Connes noncommutative geometry scenario of
gravity and electro-weak interactions and string theory.

To say simply that modular theory also occurs in Connes theory is not very
revealing since its euclidean use just amounts to the euclidean charge
conjugation and not to localization. In fact Connes recent attempts at
gravity are probably more related to replace localization by another
principle but it is presently completely unclear what this could mean in
physical terms. For people who believe in the power of analogies one should
perhaps point out that the DFR model, in which the spacetime labeling of
nets is replaced by noncommutative spacetime, leads to the same
double-sheetedness as in Connes theory. The spacetime uncertainty relations
in the DFR work are of course not characteristic, since e.g. string theory
also yields such relation. The strength and beauty of the DFR \cite{DFR}
approach lies in the fact that the authors are fully conscientious that they
are trying to tinker with the number one principle of QFT namely
localization, whose unsubtle removal would also destroy the physical
interpretability. Therefore their proposal has the very desirable feature of
being refutable by further future work. The physical alternative to
causality as needed in Quantum Gravity is not {\it no causality} or {\it %
cutoffs in certain integrals,} but rather a new principle which does not
wreck interpretation.

This brings us to a very important issue: a physical theory should carry its
own interpretation with it. Local Quantum Physics fulfills this requirement;
it has a localization concept from which one can derive scattering theory
and a wealth of formulae relating to observations. A theory which only
exists in euclidean (imaginary time) form cannot yet be called physical
because one cannot derive this wealth of formulas but only plug in some
extrapolated euclidean expressions. String theory was born as a proposal for
a crossing symmetric S-matrix. Its later use in interpreting it as a kind of
extended field theory resulted from purely formal games with the formalism
and not by a conceptual analysis which would include the important issue of
localization (or its unknown potential substitute). This leaves the
question: why does string theory have this unreasonably seeming mathematical
power if it comes just from physical formalism ``running amok'' (being the
formalism of the old dual model augmented by differential geometry and
``chased up'' to 10$^{19}$ GEV\footnote{%
As a student of Harry Lehmann I remember the following words of my adviser:
``the chance that a particular formalism implementing a physical concept
will still be valid after moving up in energy by two orders of magnitude
will be smaller than 50\%''. It seems to me that a new sociological effect
was left out: the popularity and reputation of a mathematically attractive
theory grow slowly with its decreasing experimental verificability and this
curve after sufficiently many orders of magnitude crosses the previous
decreasing curve.}.) together with geometrical interpretation instead of
physical concepts and principles? My tentative answer to this perplexing
question is that behind that mathematical success there is an yet invisible
form of the modular theory and the closely related subfactor theory. Since I
am not an expert on string theory, I can only base my arguments on facts
which have been seen in the string theoretic mode of thinking but which can
be checked in low dimensional QFT models. Among the many illustrations which
come to my mind, I will only mention one: the modular relation (in the
Gepner-Witten and Capelli-Itzykson-Zuber\cite{GWCIZ} sense of the word
``modular'') for chiral conformal correlation function which relate the
would be euclidean correlations associated to the torus and which for the
zero-point characters was proposed on geometric reasoning by Verlinde. Note
that we are avoiding to say ``on the torus'', since this would be
dangerously close to ``localized on'' which is the wrong physical picture.
These modular identities for chiral correlation functions are very profound
properties because they generalize the well known ``Nelson relations'' of
temperature correlations of box-quantized fields which is the rigorous
version of the formal symmetry of the Feynman-Kac representation under
interchange of space with euclidean time: the period caused by the
temperature may be interchanged with the size of the interval without
changing the correlations. Although my past attempts to derive the
geometrical modular identities from T.T.-modular theory failed \cite{Lo},
R.Longo was at least able to prove a weaker theorem from T.T.-modular
theory. This illustrates that by geometrical consistency arguments one is
able to discover a relation whose proper quantum interpretation (and proper
physical analogies in higher dimensions) requires the modular construction
of a (euclidean) dual euclidean theory. The physical weakness of such a
geometric consistency arguments is explained by the fact that it has not
been understood as a result of the principles of Local Quantum Physics.

Returning to Alain Connes method of noncommutative geometry, one realizes
that there is a very surprising, almost philosophical connection with some
of its recent mathematical output \cite{Connes}: the impressive
distinguished role of type III hyperfinite von Neumann factors in Connes
extended Galois theory and penetrating study of the Riemann $\zeta $%
-function. In order to see this from the physical side, let us remind
ourselves of the ``philosophical'' underpinnings of algebraic QFT as Rudolf
Haag expressed them at one occasion. The standard most widely accepted
picture of physical reality is that of Newton and Einstein: a space-time
manifold with a material content. On the other hand Leibnitz picture was
different: ``Monades'' which have no individuality by themselves but which
create reality by their interrelations. This fits precisely the nets of
observables of algebraic QFT where a monad corresponds to the hyperfinite $%
III_{1}$ factor (``if you have seen one, you know them all''). Whether the
fundamental appearance of type III factors in Connes extended Galois theory
harmonizes with the same philosophy about physical reality is certainly food
for future thoughts. In any case, the concepts and formalism developing
around modular localization constitute a change of paradigm in QFT, and
should be critically confronted in any attempt to go beyond ``Laboratory
QFT'' into the direction of Quantum Gravity.

As already mentioned in reference \cite{S2}, this work is appearantly
related to that of Max Niedermaier \cite{hep}. What seems to be in common is
the recognition of the thermal aspect of the factorization program, but a
detailed comparision is still a task for the future.

Acknowledgements:

I am indebted to Henning Rehren and Raymond Stora for critical reading and
several suggestions which led to improvements of the manuscript. My thanks
go to the organizers of the ESI ``Workshop on Algebraic QFT'' for the
invitation and the hospitality during my visit.

\end{document}